\documentclass[superscriptaddress,showkeys,showpacs,12pt]{revtex4}
\usepackage{graphicx}
\usepackage{amsmath}
\usepackage{booktabs}
\usepackage{amssymb}
\usepackage{multirow}
\usepackage{makecell}
\usepackage{graphicx}
\usepackage{textcomp}
\usepackage{subfigure}
\usepackage{phonetic}
\usepackage{extarrows}
\usepackage{color}
\usepackage{float}
\usepackage[colorlinks,citecolor=blue,linkcolor=blue]{hyperref}
\begin{document}
\title{Manipulation of nanomechanical resonator via shaking optical frequency}
\author{Dong-Yang Wang}
\affiliation{Department of Physics, Harbin Institute of Technology, Harbin, Heilongjiang 150001, China}
\author{Cheng-Hua Bai}
\affiliation{Department of Physics, Harbin Institute of Technology, Harbin, Heilongjiang 150001, China}
\author{Shutian Liu\footnote{E-mail: stliu@hit.edu.cn}}
\affiliation{Department of Physics, Harbin Institute of Technology, Harbin, Heilongjiang 150001, China}
\author{Shou Zhang\footnote{E-mail: szhang@ybu.edu.cn}}
\affiliation{Department of Physics, College of Science, Yanbian University, Yanji, Jilin 133002, China}
\author{Hong-Fu Wang\footnote{E-mail: hfwang@ybu.edu.cn}}
\affiliation{Department of Physics, College of Science, Yanbian University, Yanji, Jilin 133002, China}

\begin{abstract}
In the usual optomechanical systems, the stability of the systems severely limits those researches of the macroscopic quantum effects. We study an usual cavity optomechanical system where the frequency of the optical mode is shaken periodically. We find that, when the optical shaking frequency is large enough, the shake of the optical mode can stabilize the system. That means we can study the macroscopic quantum effects of the mechanical resonator even in the strong coupling region where the standard optomechanical systems are always unstable. As examples, we study the ground-state cooling of the mechanical resonator and the entanglement between the optical and mechanical modes in the conventional unstable region, and the results indicate that the final mean phonon number and entanglement not only can be achieved but also can be modulated by the optical shaking parameters. Our proposal provides a method to study the macroscopic quantum effects even in conventional unstable region.
\end{abstract}
\maketitle

\section{Introduction}\label{sec.1}
Cavity optomechanical system has made rapid advances in the past decades, which is mainly used to study the macroscopic quantum effects of the micromechanical resonators, such as the ground-state cooling of the mechanical resonator~\cite{PhysRevLett.80.688,Wilson2015,PhysRevLett.99.093901,2008NJP.10.095007,PhysRevLett.99.093902,PhysRevLett.110.153606,nature.541.191,PhysRevLett.119.123603}, mechanical squeezing~\cite{Wollman952,wang16SR,PhysRevA.91.013834}, entanglement~\cite{PhysRevLett.98.030405,PhysRevLett.99.250401,bai17SR,2018NJP.20.083004}, macroscopic quantum superposition~\cite{PhysRevLett.116.163602}, etc. Owing to the unique advantages of optomechanical systems, numerous potential applications have been proposed, e.g., the ultrahigh precision metrology~\cite{RevModPhys.52.341,RevModPhys.68.755}, exploring the quantum-classical-mechanics boundary~\cite{PhysRevLett.88.120401,PhysRevLett.97.237201}, and studying the weak signal transducer. The progress of the gravitational-wave detection is a great example for the application of optomechanical systems~\cite{PhysRevLett.116.061102}. 

In recent years, the periodically modulated optomechanical systems have attracted significant attention, which have been used to study various macroscopic quantum effects~\cite{PhysRevLett.103.213603,PhysRevA.83.033820,PhysRevA.86.013820,PhysRevA.89.023843,PhysRevA.95.053861}. However, in those modulation proposals, most of them focus on modulating the driven laser field, which results in the first-order moments of the system operators and the effective optomechanical coupling changing periodically to achieve and study some quantum effects. On the other hand, the frequency modulated quantum systems also exhibit a rich behavior and display nonequilibrium properties that are absent in their static counterparts~\cite{PhysRevA.96.043849,Rep.Prog.Phys.80.056002}, such as the phenomena of motional averaging and narrowing~\cite{doi:10.1143/JPSJ.9.316}, Landau--Zener--St\"{u}ckelberg--Majorana interference~\cite{SHEVCHENKO20101}, and the formation of dressed states with the appearance of sidebands in the spectrum. However, in cavity optomechanical systems, the study of the influence coming from the frequency modulation is relatively rare to date~\cite{2018NJP.20.083024,PhysRevA.92.013822,PhysRevA.98.023816,2016JO.18.084001}. In Ref.~\cite{PhysRevA.92.013822}, the authors enhance the mechanical effects of single photons via introducing a resonance-frequency modulation to the two cavity fields in the cavity--membrane--cavity optomechanical system. We have studied the improved ground-state cooling of the mechanical resonator via modulating both the optical and mechanical frequencies simultaneously~\cite{PhysRevA.98.023816}. In addition, most of the previous researches about the macroscopic quantum effects are limited in the stable region of the usual optomechanical systems, which dependents on the system parameters, especially the effective optomechanical coupling strength. So how to study the macroscopic quantum effects in unstable regions is of momentous significance. As far as we know, the approach of only modulating the optical frequency to study the steady macroscopic quantum effects in unstable region has not yet been reported.

In this paper, we study an usual cavity optomechanical system where the frequency of the optical mode is modulated periodically. As we all know, the stability of optomechanical systems is closely related to the effective optomechanical coupling strength. For an excessively large coupling strength, the optomechanical systems are unstable and the studying is also meaningless. However, we find that the shaking optical mode can reduce the effective optomechanical coupling strength arbitrarily when the shaking frequency is much larger than the mechanical resonator frequency, and the deeply physical mechanism can be explained by the Raman-scattering or frequency-domain pictures. The result indicates that it will be possible to study the steady quantum effects of optomechanical system even with strong coupling where the standard optomechanical systems without frequency modulation are always unstable. In order to verify the above analyses, we study the ground-state cooling of the mechanical resonator and the entanglement between the optical and mechanical modes when the effective optomechanical coupling belongs to the conventional unstable region of the standard optomechanical systems. Moreover, we also find that the final mean phonon number and the steady entanglement are related to the shaking parameters of the optical frequency, which indicates that we can manipulate the quantum effects of the mechanical resonator via changing the optical shaking parameters.

The paper is organized as follows: In Sec.~\ref{sec.2}, we derive the linearized Hamiltonian of the optomechanical system with frequency modulation and explain the physical mechanism through the Raman-scattering and frequency domain pictures. In Sec.~\ref{sec.3}, we give the correlation matrix of the system operators. In Sec.~\ref{sec.4}, we first study the ground-state cooling of the mechanical resonator with strong optomechanical coupling in the presence of the frequency modulation and discuss the effect of modulation parameters on the final mean phonon number. In Sec.~\ref{sec.5}, we calculate the entanglement between the optical and mechanical modes by utilizing the logarithmic negativity and give the influence of different system parameters, respectively. Finally, a conclusion is given in Sec.~\ref{sec.6}.

\section{System and Hamiltonian}\label{sec.2}
\begin{figure}
	\centering
	\includegraphics[width=1.0\linewidth]{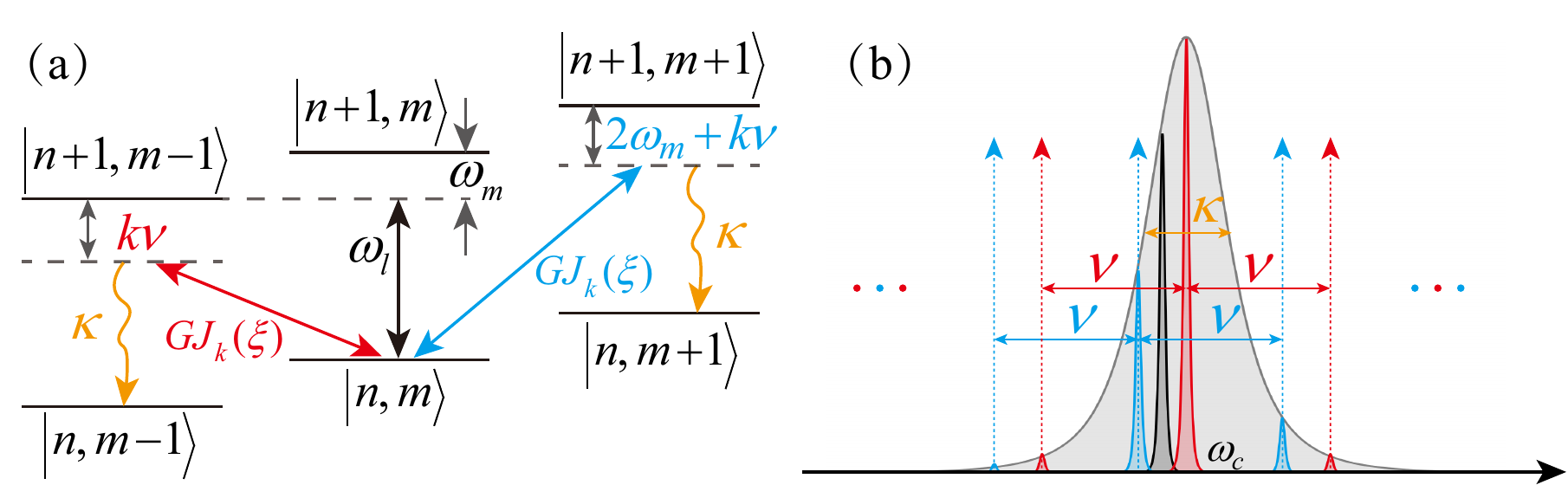}
	\caption{(a) Level diagram of the linearized Hamiltonian with frequency modulation [Eq.~(\ref{e04})]. $|n, m\rangle$ denotes the state of $n$ photons and $m$ phonons in the displaced frame, the red (blue) arrow represents the beam-splitter (two-mode squeezing) interactions, the black arrow represents the driven laser, and the yellow arrows represent the leakage of optical mode. (b) Frequency-domain interpretation of the optomechanical interactions in the presence of frequency modulation.}
	\label{fig:Red-Raman}
\end{figure}

We consider an usual optomechanical system\textemdash including one mechanical resonator and one optical mode\textemdash in which the frequency of optical mode is cosine modulated. In the rotating frame at the driven laser frequency $\omega_{l}$, the Hamiltonian of the system is given as ($\hbar=1$)
\begin{eqnarray}\label{e01}
\widetilde{H}=\left[\Delta_{c}+\xi\nu\cos(\nu t)\right]a^{\dagger}a+\omega_{m}b^{\dagger}b
  -ga^{\dagger}a(b^{\dagger}+b)+(Ea^{\dagger}+E^{\ast}a),
\end{eqnarray}
where $a~(b)$ and $a^{\dagger}~(b^{\dagger})$ represent the annihilation and creation operators for optical (mechanical) mode with frequency $\omega_{c}~(\omega_{m})$, respectively. $\Delta_{c}=\omega_{c}-\omega_{l}$ is the original cavity laser detuning, $\xi$ is the normalized modulation amplitude, $\nu$ is the modulation frequency, $g$ is the single photon optomechanical coupling strength, and $E=\sqrt{2\kappa P/(\hbar\omega_{l})}$ is the amplitude of the driven laser, where $\kappa$ is the decay rate of the optical mode and $P$ is the power of the driven laser. In the presence of frequency modulation, using the usual linearization approach (see Appendix~\ref{App1}), e.g., $a=\alpha+\delta a$ and $b=\beta+\delta b$, we can derive the linearized Hamiltonian
\begin{eqnarray}\label{e02}
H_{L}=\left[\Delta_{c}^{\prime}+\xi\nu\cos(\nu t)\right]\delta a^{\dagger}\delta a+\omega_{m}\delta b^{\dagger}\delta b
	-(G\delta a^{\dagger}+G^{\ast}\delta a)(\delta b^{\dagger}+\delta b),
\end{eqnarray}
where $\Delta_{c}^{\prime}=\Delta_{c}-g(\beta+\beta^{\ast})$ and $G=g\alpha$ is linearized optomechanical coupling strength. To show the effect from shaking the optical frequency clearly, we perform the rotating transformation defined by
\begin{eqnarray}\label{e03}
V_{2}&=&\mathcal{T}\exp\left\{-i\int_{0}^{t}d\tau\left[\Delta_{c}^{\prime}
+\xi\nu\cos(\nu t)\right]\delta a^{\dagger}\delta a+\omega_{m}\delta b^{\dagger}\delta b\right\}\cr\cr
&=&\exp\left\{-i\left[\Delta_{c}^{\prime}t+\xi\sin(\nu t)\right]\delta a^{\dagger}\delta a
-i\omega_{m}t\delta b^{\dagger}\delta b\right\},
\end{eqnarray}
where $\mathcal{T}$ denotes the time ordering operator. In the rotating frame defined by the transformation operator $V_{2}$, the transformed Hamiltonian is derived as
\begin{eqnarray}\label{e04}
\widetilde{H}_{\mathrm{LM}}&=&V_{2}^{\dag}H_{\mathrm{LM}}V_{2}-iV_{2}^{\dag}\dot{V}_{2}\cr\cr
				&=&-G\left(\delta a^{\dagger}\delta b^{\dagger}e^{i[(\Delta_{c}^{\prime}+\omega_{m})t+\xi\sin(\nu t)]}
				+\delta a^{\dagger}\delta be^{i[(\Delta_{c}^{\prime}-\omega_{m})t+\xi\sin(\nu t)]}\right)+\mathrm{H.c.}\cr\cr
				&=&-\sum_{k=-\infty}^{\infty}\left[GJ_{k}(\xi)\delta a^{\dagger}\delta b^{\dagger}e^{i(\Delta_{c}^{\prime}+\omega_{m}+k\nu)t}
				+GJ_{k}(\xi)\delta a^{\dagger}\delta be^{i(\Delta_{c}^{\prime}-\omega_{m}+k\nu)t}+\mathrm{H.c.}\right].
\end{eqnarray}
The above derivation needs the Jacobi--Anger expansions: $e^{i\xi\sin(\nu t)}=\sum_{k=-\infty}^{\infty}J_{k}(\xi)e^{ik\nu t}$, where $J_{k}(\xi)$ is the Bessel function of the first kind with $k$ being an integer. The physics process of Eq.~(\ref{e04}) can be explained via the Raman-scattering picture Fig.~\ref{fig:Red-Raman}(a) or the frequency-domain picture Fig.~\ref{fig:Red-Raman}(b). Under the red detuning sideband resonant condition ($\Delta_{c}^{\prime}=\omega_{m}$), $|n, m\rangle$ denotes the state of $n$ photons and $m$ phonons in the displaced frame, the red arrow represents the beam-splitter interaction with the coupling strength $GJ_{k}(\xi)$ and detuning $k\nu$, the blue arrow represents the two-mode squeezing interaction with coupling strength $GJ_{k}(\xi)$ and detuning $2\omega_{m}+k\nu$, the black arrow represents the driven laser, and the yellow arrows represent the leakage of the optical mode. In Fig.~\ref{fig:Red-Raman}(b), the red (blue) peaks represent the beam-splitter (two-mode squeezing) interactions corresponding to different sidebands, respectively, which are discrete and have been separated with modulation frequency $\nu$ space. We can see that the nearest resonant sideband ($k=k_{0}$) of all is the leading order in optomechanical interactions and the effective coupling strength [$GJ_{k_{0}}(\xi)$] is modulated by parameter $\xi$, independently. Moreover, if the modulation frequency $\nu$ is large enough, the other sidebands will be negligible due to far from the resonator condition. Then the Hamiltonian~(\ref{e04}) can be reduced to the standard optomechanical Hamiltonian with effective coupling $GJ_{k_{0}}(\xi)$, which is modified by the Bessel function of the first kind $J_{k_{0}}(\xi)$ and infers that we can study the quantum effects of the optomechanical system via changing the optical shaking parameters.

\section{Correlation matrix of the system}\label{sec.3}
\begin{figure}
	\centering
	\includegraphics[width=0.49\linewidth]{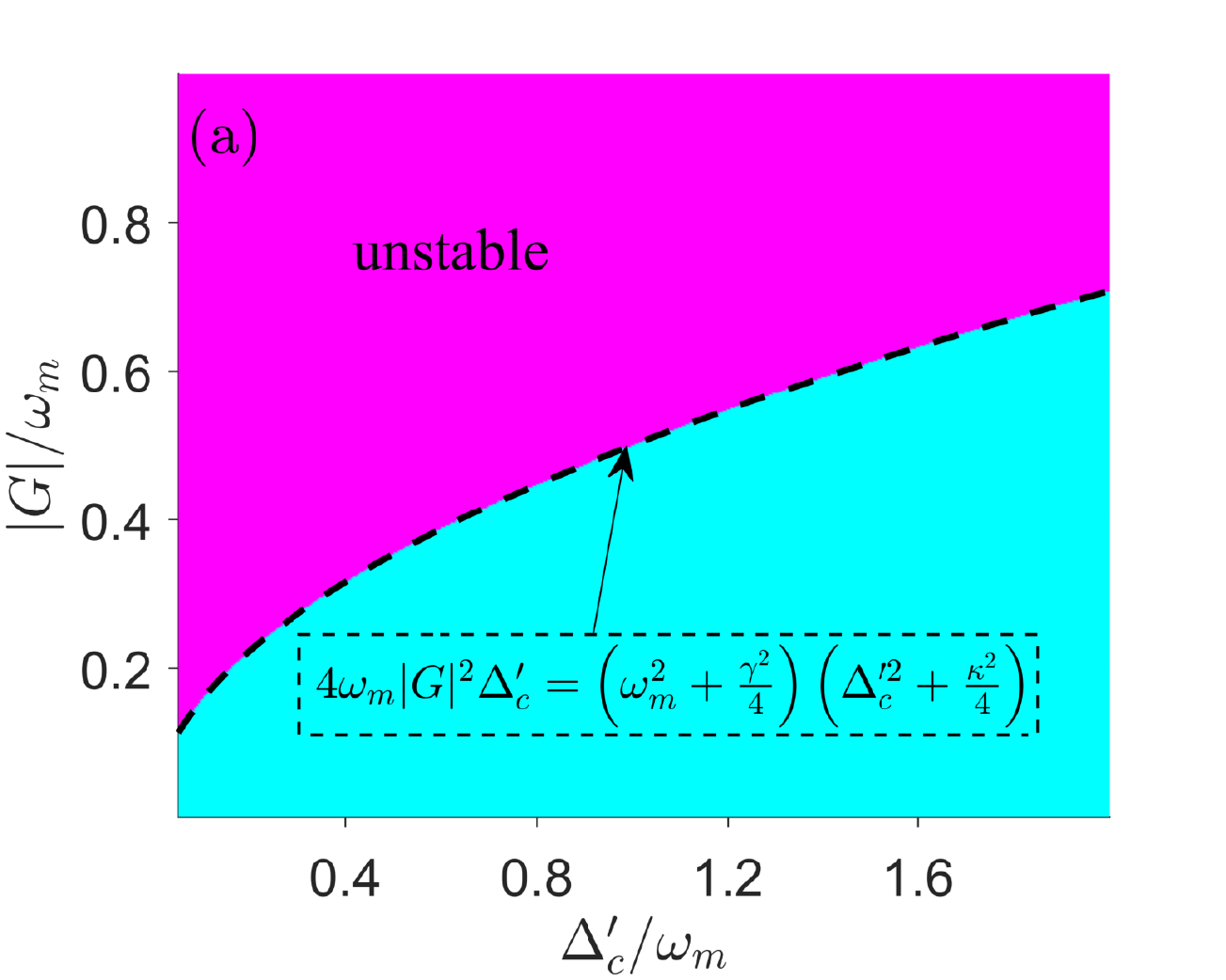}
	\hspace{0in}%
	\includegraphics[width=0.495\linewidth]{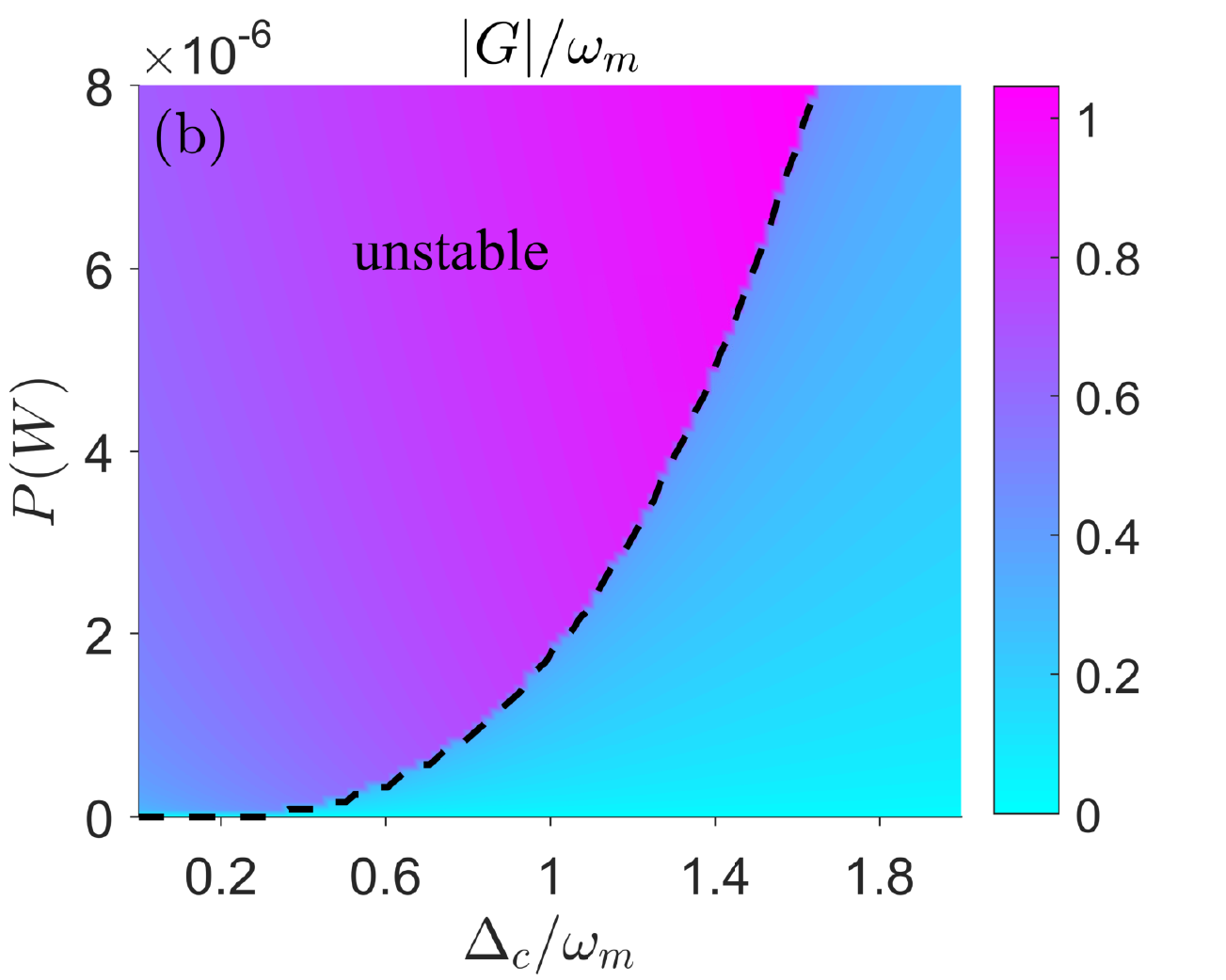}
	\caption{The stability of the conventional optomechanical system without frequency modulation. (a) The stability relates to the effective detuning $\Delta_{c}^{\prime}$ and effective optomechanical coupling $G$. The black dashed line is the boundary of stable region, which can be derived via Routh--Hurwitz criterion~\cite{PhysRevA.35.5288}. (b) The effective optomechanical coupling relates to the cavity laser detuning $\Delta_{c}$ and power of the driven laser. The black dashed line is the boundary of stable region determined by numerically solving the eigenvalues of Eq.~(\ref{e05}). The parameters are chosen as: $\omega_{c}=2\pi\times7.54~\mathrm{GHz}$, $\kappa=2\pi\times200~\mathrm{kHz}$, $\omega_{m}=2\pi\times10.56~\mathrm{MHz}$, $\gamma=2\pi\times32~\mathrm{Hz}$, and $g=2\pi\times200~\mathrm{Hz}$, which come from Ref.~\cite{nature10261}.}
	\label{fig:stability}
\end{figure}

To study the dynamics and quantum effects of the system, we utilize the correlation matrix of the system operators as main processing method. Firstly, it is convenient to define the quadrature components of optical and mechanical modes, i.e., $x=(\delta a^{\dagger}+\delta a)/\sqrt{2}$, $y=i(\delta a^{\dagger}-\delta a)/\sqrt{2}$, $q=(\delta b^{\dagger}+\delta b)/\sqrt{2}$, and $p=i(\delta b^{\dagger}-\delta b)/\sqrt{2}$. Similarly, the corresponding noise quadratures are given by $x_{\mathrm{in}}=(a_{\mathrm{in}}^{\dagger}+a_{\mathrm{in}})/\sqrt{2}$, $y_{\mathrm{in}}=i(a_{\mathrm{in}}^{\dagger}-a_{\mathrm{in}})/\sqrt{2}$, $q_{\mathrm{in}}=(b_{\mathrm{in}}^{\dagger}+b_{\mathrm{in}})/\sqrt{2}$, and $p_{\mathrm{in}}=i(b_{\mathrm{in}}^{\dagger}-b_{\mathrm{in}})/\sqrt{2}$. Based on Eq.~(\ref{Ae04}), we derive and give the dynamical equation of those quadrature components in the compact form, i.e., $\dot{u}=Au-n$, where $u=[x,y,q,p]^{T}$ represents the vector of quadrature components, $n=[\sqrt{\kappa}x_{\mathrm{in}},\sqrt{\kappa}y_{\mathrm{in}},\sqrt{\gamma}q_{\mathrm{in}},\sqrt{\gamma}p_{\mathrm{in}}]^{T}$ represents the vector of noise quadratures, and $A$ is a $4\times4$ matrix given by
\begin{eqnarray}\label{e05}
A=\left(\begin{array}{cccc}
~~~-\frac{\kappa}{2}~~~&~~~\Delta~~~&~~~-2\mathrm{Im}[G]~~~&~~~0~~~\\ 
-\Delta & -\frac{\kappa}{2} & 2\mathrm{Re}[G] & 0 \\ 
0 & 0 & -\frac{\gamma}{2} & \omega_{m} \\ 
2\mathrm{Re}[G]	& 2\mathrm{Im}[G] & -\omega_{m} & -\frac{\gamma}{2}
\end{array}\right),
\end{eqnarray}
where $\Delta=\Delta_{c}^{\prime}+\xi\nu\cos(\nu t)$ is the effective detuning. For the conventional optomechanical system without frequency modulation ($\xi=0$), the stability is determined through Routh--Hurwitz criterion, which is a mathematical test for the stability of a linear time-invariant system, and the result is shown in Fig.~\ref{fig:stability}. We can see that the conventional optomechanical system is unstable for too large optomechanical coupling or too large power of the driven laser. However, it is worth noting that the system with frequency modulation changes to time-dependent system so that the criterion is no longer applicable. So we determine the stability of the system with frequency modulation via observing the dynamical behavior of the system in our work.

Additionally, the above dynamical equation of quadrature components is related to the quantum fluctuations part, which is stochastic and unsuited to directly describe the quantum effects of the system. Thanks to the zero-mean Gaussian nature and autocorrelation functions given by Eq.~(\ref{Ae05}) for the quantum noises, we can introduce the correlation matrix $V$ of quadrature components to describe the quantum effects of the system. The corresponding matrix element is defined by $V_{ij}=\langle u_{i}u_{j}+u_{j}u_{i}\rangle/2$. Then we can derive the motion equation for the correlation matrix as
\begin{eqnarray}\label{e06}
\dot{V}=AV+VA^{T}+D,
\end{eqnarray}
where the diagonal diffusion matrix $D$ is related to the noise correlations and given by $D=\mathrm{diag}[\kappa/2,\kappa/2,\gamma(2n_{\mathrm{th}}+1)/2,\gamma(2n_{\mathrm{th}}+1)/2]$.

\section{Sideband cooling beyond weak coupling limit}\label{sec.4}
In the conventional optomechanical system, the ground-state cooling utilizing the red sideband ($\Delta_{c}^{\prime}=\omega_{m}$) self-cooling method cannot be achieved when the optomechanical coupling is too large~\cite{PhysRevA.80.033821,PhysRevA.83.013816,NJP.14.095015,PhysRevA.89.053821,OE.26.6143}, where the system is unstable (see Fig.~\ref{fig:stability}) and the dynamical evolution of the mean phonon number is divergent. Such as, the threshold of optomechanical coupling strength is almost $|G|=0.5\omega_{m}$ when the system is satisfied with the red sideband resonant condition $\Delta_{c}^{\prime}=\omega_{m}$. It indicates that the system is unstable when the optomechanical coupling strength is larger than $0.5\omega_{m}$ at this time. However, in the presence of frequency modulation, we prove that the ground-state cooling is a possible task even with arbitrarily large optomechanical coupling and the final mean phonon number is related to the parameters of shaking optical mode.

\begin{figure}
	\centering
	\includegraphics[width=0.49\linewidth]{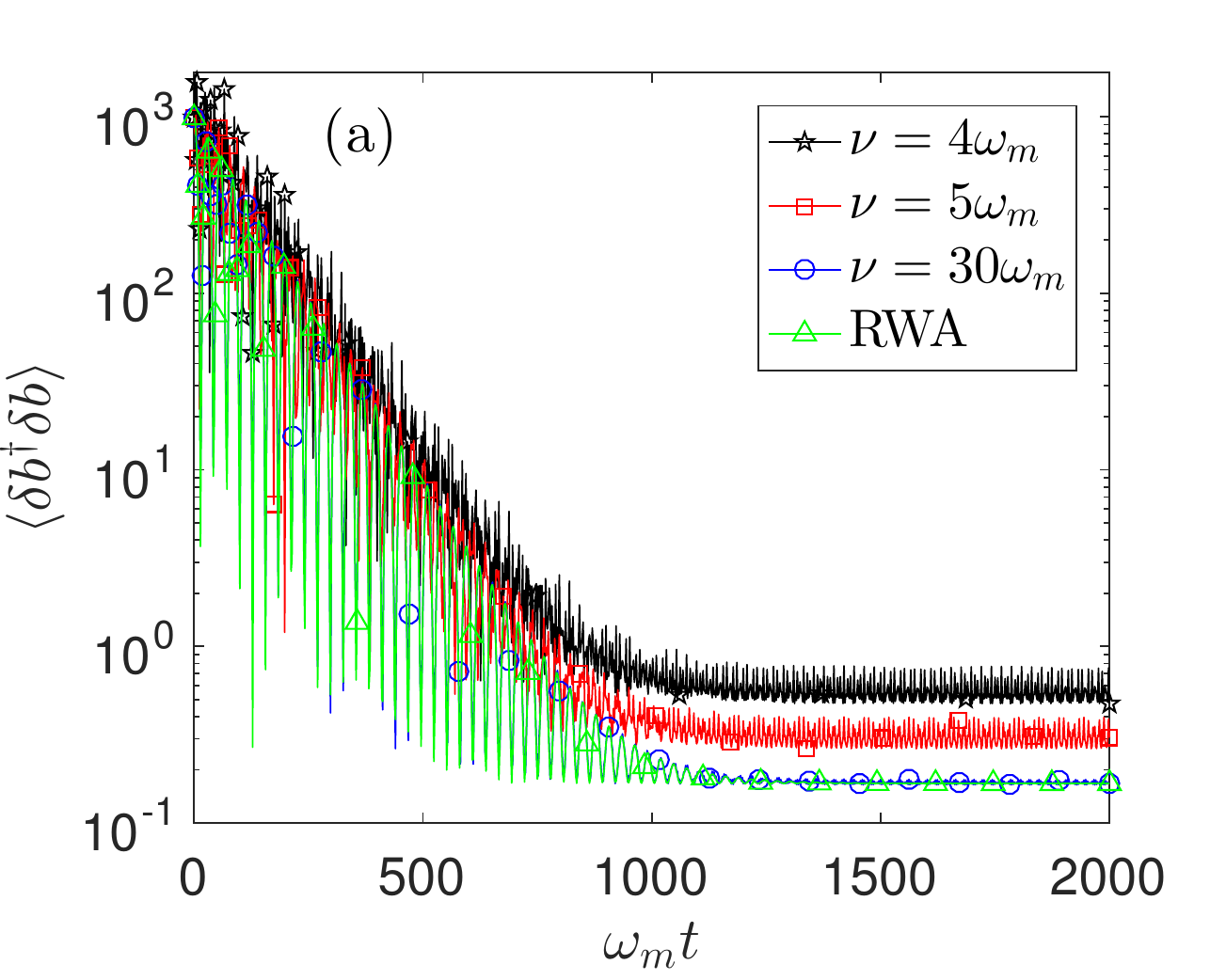}
	\hspace{0in}%
	\includegraphics[width=0.49\linewidth]{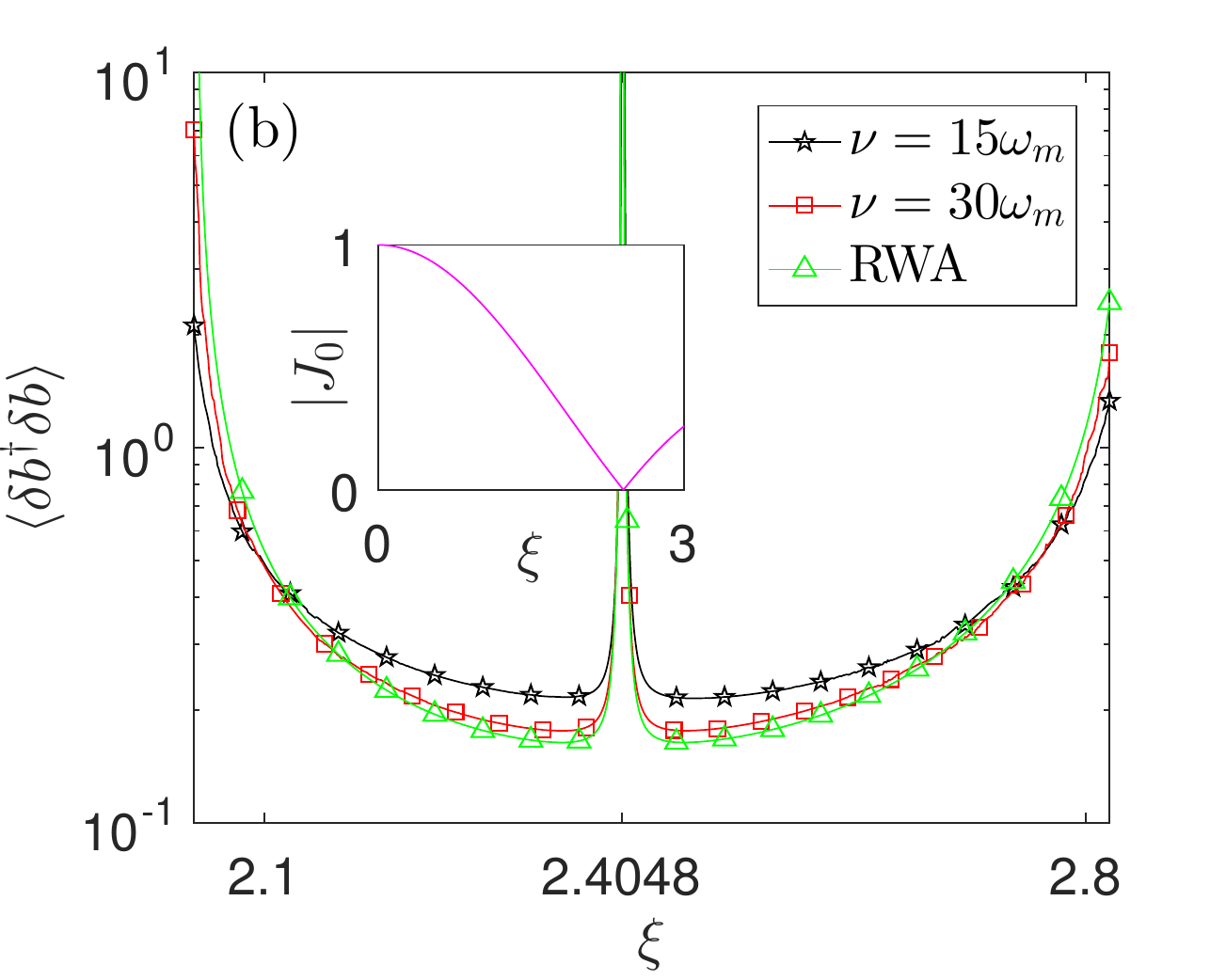}
	\caption{The final mean phonon number versus the evolution time and modulation amplitude. (a) The time evolution of mean phonon number $\langle\delta b^{\dagger}\delta b\rangle$ by solving Eq.~(\ref{e06}) for different modulation frequencies. The other parameters are chosen as: $G=\omega_{m}$, $\Delta_{c}^{\prime}=\omega_{m}$, $\xi=2.2$ and $T=500~\mathrm{mK}(n_{\mathrm{th}}\simeq10^{3})$. (b) The average of the final steady-state mean phonon number versus the modulation amplitude for different modulation frequencies. The all parameter are same as before excepting $G=2.5\omega_{m}$. The insert denotes the Bessel function $|J_{0}(\xi)|$ varying with $\xi$.}
	\label{fig:phonon-num}
\end{figure}

To explain the physical mechanism of shaking optical mode in mathematics, we simplify Eq.~(\ref{e04}) by using the rotating wave approximation (RWA). For large modulation frequency $\nu\gg\{\omega_{m},GJ_{k}(\xi)\}$, the RWA method is feasible for the Hamiltonian in Eq.~(\ref{e04}). Ignoring those sidebands with large detuning and returning to the original rotating frame, the reduced Hamiltonian is rewritten as
\begin{eqnarray}\label{e07}
H_{\mathrm{RWA}}&=&\Delta_{c}^{\prime}\delta a^{\dagger}\delta a+\omega_{m}\delta b^{\dagger}\delta b
-GJ_{0}(\xi)\left(\delta a^{\dagger}\delta b^{\dagger}+\delta a^{\dagger}\delta b\right)+\mathrm{H.c.},
\end{eqnarray}
which is the same to the conventional optomechanical Hamiltonian excepting the modified coupling strength $GJ_{0}(\xi)$. Based on the property of the Bessel function [see the insert in Fig.~\ref{fig:phonon-num}(b)], we find that the effective optomechanical coupling $GJ_{0}(\xi)$ can be arbitrarily reduced by choosing different $\xi$, which indicates that the ground-state cooling can be achieved even in arbitrarily strong coupling region. The final mean phonon number can be calculated through the mathematical expression
\begin{eqnarray}\label{e08}
\langle \delta b^{\dagger}\delta b\rangle=\frac{1}{2}\left(\langle q^{2}\rangle+\langle p^{2}\rangle-1\right),
\end{eqnarray}
where the result can be obtained by solving Eq.~(\ref{e06}) numerically, as shown in Fig.~\ref{fig:phonon-num}. Here, we have assumed that the mean phonon number of the initial system equals the thermal phonon number $\langle\delta b^{\dagger}\delta b\rangle(t=0)=n_{\mathrm{th}}$. It is worth noting that the system can be stable with those parameters when the frequency modulation is included. Moreover, in Fig.~\ref{fig:phonon-num}(a), we find that, even with the strong optomechanical coupling, the stable ground-state cooling can be achieved. That is because the frequency modulation transforms the system from strong coupling to effective weak coupling, where the system will be stable. In Fig.~\ref{fig:phonon-num}(b) we also study the transition efficiency for a stronger optomechanical coupling ($G=2.5\omega_{m}$). We find that there has a threshold [$GJ_{0}(\xi)<0.5\omega_{m}$] to limit the stability of the system for achieving the ground-state cooling, which is consistent with the stability criterion. Furthermore, we notice that the ground-state cooling cannot be achieved when $\xi$ is too close to 2.4048, which is the zero point of $J_{0}(\xi)$ vanishing the optomechanical coupling and results in the failure for ground-state cooling.

\section{Manipulating entanglement between the optical and mechanical modes}\label{sec.5}
The entanglement about nanomechanical resonator has been investigated widely in various systems and proposals~\cite{PhysRevLett.98.030405,PhysRevLett.99.250401,bai17SR,PhysRevA.89.023843,PhysRevA.84.052327,PhysRevA.86.042306}. However, most of those studies are all relied on the stable condition which can be indicated as the limit of the optomechanical coupling strength, as shown in Fig.~\ref{fig:stability}. Through the above analysis, we have found that the shaking optical frequency can reduce the effective coupling strength of the optomechanical system arbitrarily. In our proposed cavity optomechanical system, therefore, the manipulation of steady entanglement about macroscopic mechanical resonator can also be achieved even in the strong coupling region where the standard optomechanical systems are always unstable

In order to estimate the entanglement between the optical and mechanical modes, we utilize the logarithmic negativity $E_{N}$ as the quantity to measure the entanglement, which can be calculated via the definition
\begin{eqnarray}\label{e09}
E_{N}=\max[0,-\ln 2\eta^{-}],
\end{eqnarray}
where $\eta^{-}=2^{-1/2}\{\Sigma(V)-[\Sigma(V)^{2}-4\det V]^{1/2}\}^{1/2}$, with $\Sigma(V)=\det A+\det B-2\det C$, and the correlation matrix $V$ by the $2\times2$ block matrices is given as
\begin{eqnarray}\label{e10}
V=\left(\begin{array}{cc}
A~&~C\\ 
C^{T} & B
\end{array}\right).
\end{eqnarray}
Therefore, the optical mode and mechanical mode are said to be entangled ($E_{N}>0$) if and only if $\eta^{-}<1/2$, which is equivalent to the Simon's necessary and sufficient nonpositive partial transpose criteria~\cite{PhysRevLett.98.030405,PhysRevLett.84.2726}.

\begin{figure}
	\centering
	\includegraphics[width=0.49\linewidth]{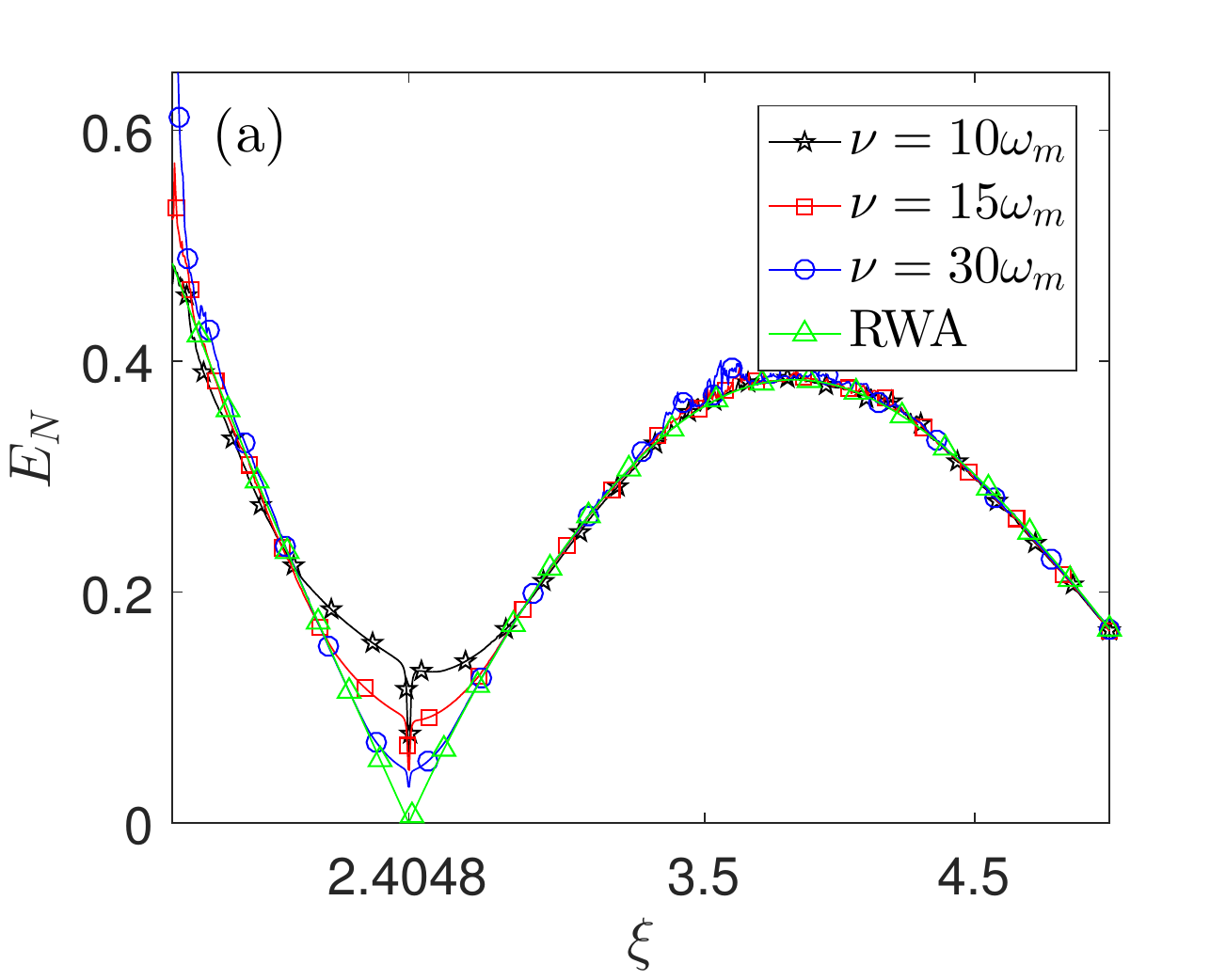}
	\hspace{0in}%
	\includegraphics[width=0.49\linewidth]{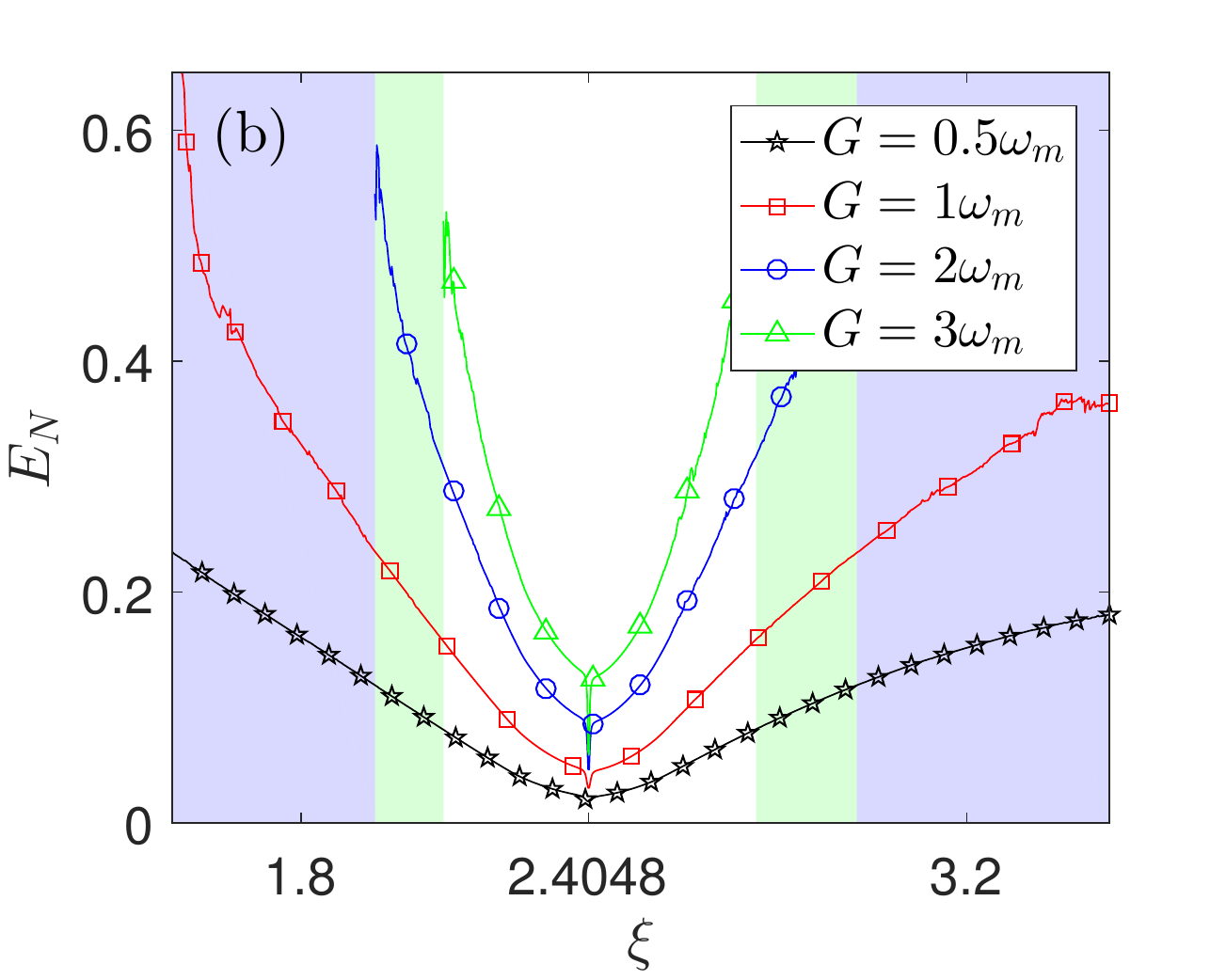}
	\vspace{0in}%
	\includegraphics[width=0.49\linewidth]{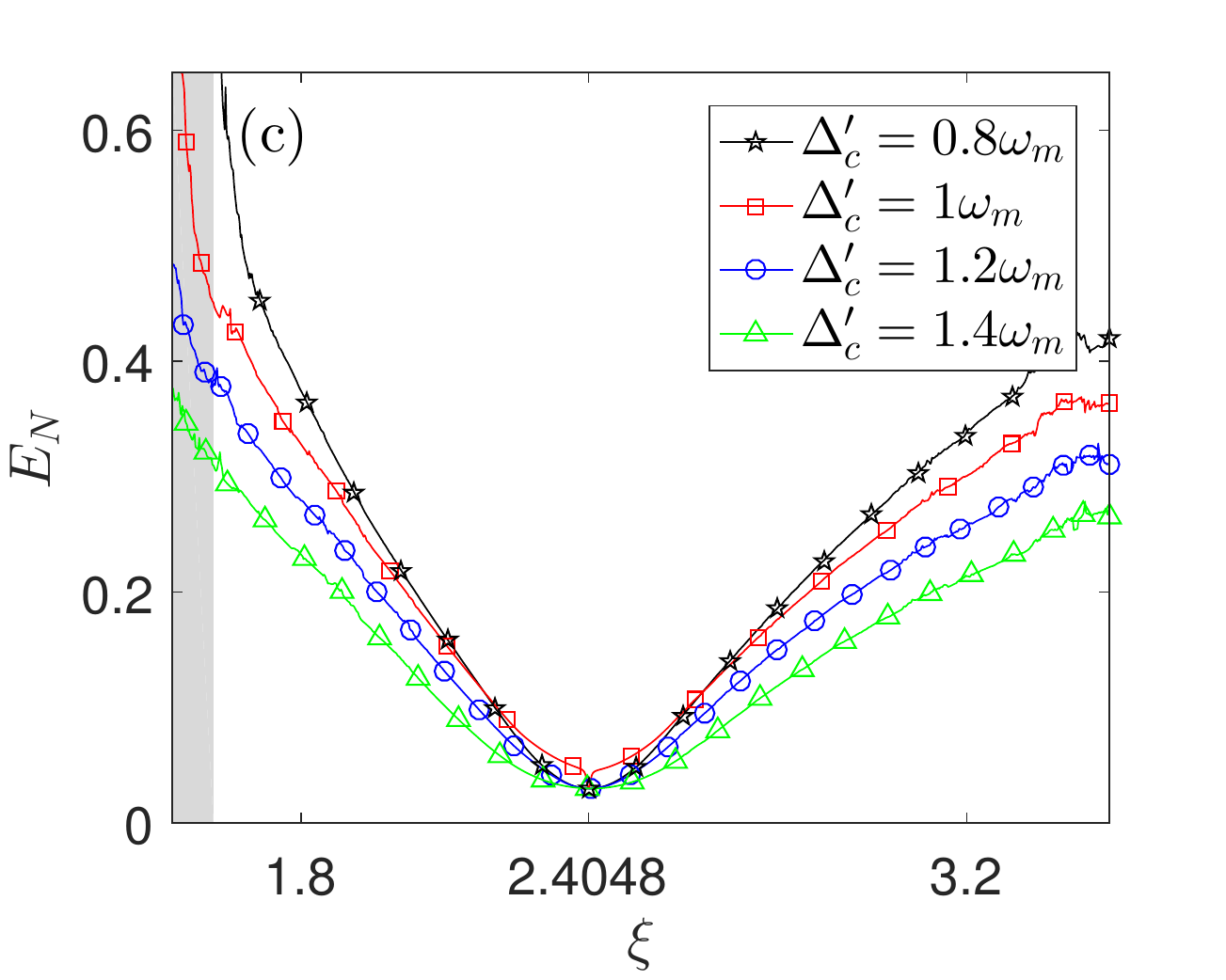}
	\hspace{0in}%
	\includegraphics[width=0.49\linewidth]{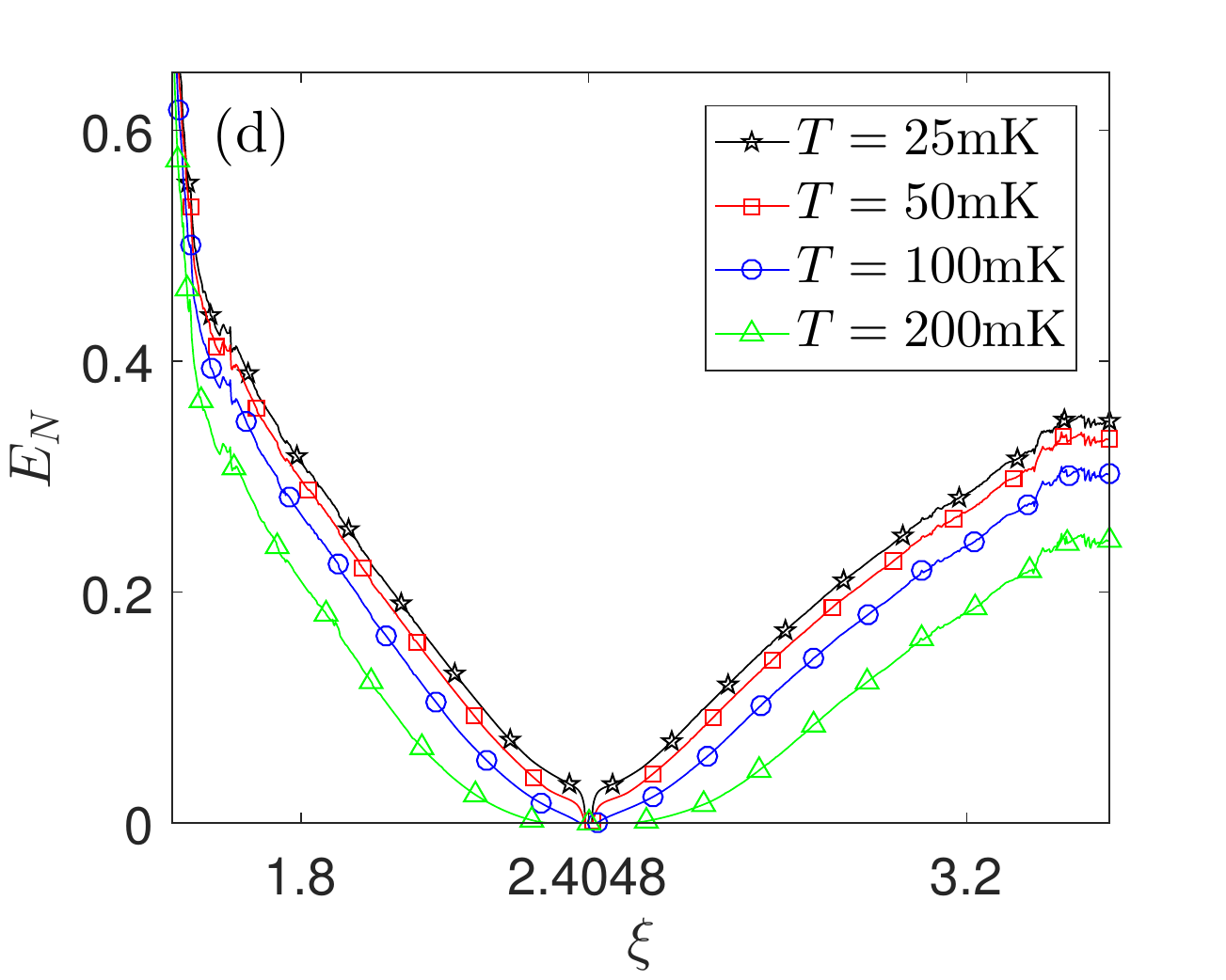}
	\caption{The average of the final steady-state entanglement $E_{N}$ versus the modulation amplitude for different system parameters. (a) Different modulation frequencies. (b) Different optomechanical couplings. (c) Different effective detunings. (d) Different temperatures. The other parameters are chosen as: $\nu=30\omega_{m}$, $G=\omega_{m}$, $\Delta_{c}^{\prime}=\omega_{m}$,  and $T=0~\mathrm{mK}$. The shadows in Fig.~(b-c) represent the corresponding unstable regions with those system parameters.}
	\label{fig:En-xi}
\end{figure}

The relationships between logarithmic negativity $E_{N}$ and modulation amplitude $\xi$ with different parameters are shown in Fig.~\ref{fig:En-xi}, e.g., corresponding to different modulation frequencies, optomechanical couplings, effective detunings, and temperatures, respectively. In Fig.~\ref{fig:En-xi}(a), we first show $E_{N}$ changing with $\xi$ for different modulation frequencies. One can see that $E_{N}$ has a sharp decline at the zero point of $J_{0}(\xi)$ due to the too weak coupling of the nearest resonant sideband, but it does not vanish due to the existence of those non-RWA terms (other sidebands). Then we study the change of entanglement with different optomechanical couplings and show the result in Fig.~\ref{fig:En-xi}(b). We find that the system with frequency modulation is stable and the entanglement is existent even in the conventional unstable region of the standard optomechanical systems. However, the adjustable range of $\xi$ decreases gradually with the increase of optomechanical coupling $G$. That is because the larger coupling needs the smaller $J_{0}(\xi)$ to ensure the system being stable. Finally, the effects of different effective detunings and temperatures are also studied respectively, as shown in Fig.~\ref{fig:En-xi}(c-d). We find that the entanglement always exists near $\Delta_{c}^{\prime}=\omega_{m}$ and changes with the parameter $\xi$. Moreover, the entanglement $E_{N}$ decreases with the increase of the bath temperature, and it even disappears when the effective coupling is too weak, which is consistent with the previous studies. Overall, we study the steady entanglement between the optical and mechanical modes in an optomechanical system with frequency modulation when the optomechanical coupling $G$ is very lager, where the standard optomechanical systems are unstable. We find that the system with frequency modulation can be stable and the entanglement exists even in conventional unstable region, and the entanglement can also be manipulated by the parameters of shaking optical mode.

Finally, we briefly discuss the implementation of the present proposal. The usual optomechanical systems have been studied and realized experimentally~\cite{Nat.Phys.4.415,Nature.463.72,Nature.556.478}, and the experimental detection of those quantum effects about the micromechanical resonator has been reported~\cite{nphys.12.683,nature16536}. The innovative technology requested in our proposal is the frequency modulation on the optical mode, which has also been reported~\cite{Rep.Prog.Phys.80.056002,PhysRevLett.102.090501}. In our proposal, the modulation frequency on optical mode should be several times as much as the mechanical frequency, which is much smaller than the optical frequency in the usual optomechanical systems. Therefore, the proposed proposal is feasible in experiment with current experimental techniques.

\section{Conclusions}\label{sec.6}
In conclusion, we have studied an usual cavity optomechanical system where the frequency of the optical cavity is modulated in the form of cosine. By using the usual linearization approach and transforming the system to the interaction picture, we explain the physical mechanism via the Raman-scattering and frequency-domain pictures. We find that the frequency modulation of the optical mode can reduce the optomechanical coupling strength arbitrarily when the modulation frequency is large enough. Therefore, it is possible to study the steady macroscopic quantum phenomenon even in the strong coupling region where the standard cavity optomechanical system is usually unstable. We first study the ground-state cooling of the mechanical resonator by numerically solving the correlation matrix of the system and discuss the effect of modulation parameters on the final mean phonon number. The result shows that, in the presence of the frequency modulation, the mechanical ground-state cooling is not limited to the conventional stability boundary (small optomechanical coupling strength) and the final mean phonon number can also be manipulated by the modulation parameters. In addition, we also study the effect of modulation parameters on the steady entanglement between the optical and mechanical modes which is estimated by the logarithmic negativity. We find that, in the presence of frequency modulation, the steady entanglement is existent even with the very large optomechanical coupling. Our proposal would open up the possibility for studying and manipulating the macroscopic quantum effects in the strong coupling region and not limited by the stability of the system.

\begin{center}
{\bf{ACKNOWLEDGMENTS}}
\end{center}
This work was supported by the National Natural Science Foundation of China under Grant Nos.
61822114, 61465013, 61575055, and 11465020, and the Project of Jilin Science and Technology Development for Leading Talent of Science and Technology Innovation in Middle and Young and Team Project under Grant No. 20160519022JH.

\appendix
\section{Linearizing the system Hamiltonian in the presence of frequency modulation}\label{App1}
In the presence of frequency modulation, the system Hamiltonian of the standard cavity optomechanical system reads
\begin{eqnarray}\label{Ae01}
H=\left[\omega_{c}+\xi\nu\cos(\nu t)\right]a^{\dagger}a+\omega_{m}b^{\dagger}b-ga^{\dagger}a(b^{\dagger}+b)
	+(Ea^{\dagger}e^{-i\omega_{l}t}+E^{\ast}ae^{i\omega_{l}t}),
\end{eqnarray}
where $a~(b)$ and $a^{\dagger}~(b^{\dagger})$ represent the annihilation and creation operators for optical (mechanical) mode with the corresponding frequency $\omega_{c}~(\omega_{m})$, respectively. The parameter $g$ is the single photon optomechanical coupling rate. The parameters $E$ and $\omega_{l}$ are the driving amplitude and frequency, respectively. By performing a rotating transformation defined by $V_{1}=\exp[-i\omega_{l}ta^{\dagger}a]$, the transformed Hamiltonian $\widetilde{H}=V_{1}^{\dag}HV_{1}-iV_{1}^{\dag}\dot{V}_{1}$ becomes
\begin{eqnarray}\label{Ae02}
\widetilde{H}=\left[\Delta_{c}+\xi\nu\cos(\nu t)\right]a^{\dagger}a+\omega_{m}b^{\dagger}b
	-ga^{\dagger}a(b^{\dagger}+b)+(Ea^{\dagger}+E^{\ast}a),
\end{eqnarray}
where $\Delta_{c}=\omega_{c}-\omega_{l}$ is the cavity laser detuning. The quantum Langevin equations of the system are given by
\begin{eqnarray}\label{Ae03}
\dot{a}&=&-i\left[\Delta_{c}+\xi\nu\cos(\nu t)\right]a-\frac{\kappa}{2}a+iga(b^{\dagger}+b)-iE-\sqrt{\kappa}a_{\mathrm{in}},\cr\cr
\dot{b}&=&-i\omega_{m}b-\frac{\gamma}{2}b+iga^{\dagger}a-\sqrt{\gamma}b_{\mathrm{in}},
\end{eqnarray}
where $\kappa$ and $\gamma$ are the decay rate of optical cavity and the damping rate of mechanical resonator, respectively. $a_{\mathrm{in}}$ and $b_{\mathrm{in}}$ are the corresponding noise operators. Under the strongly coherent laser driving, we can apply a displacement transformation to linearize  Eq.~(\ref{Ae03}), i.e., $a=\alpha+\delta a$ and $b=\beta+\delta b$, where $\alpha$ and $\beta$ are $c$-numbers representing the displacement mean values of the cavity and mechanical resonator modes. $\delta a$ and $\delta b$ are the operators relating to the quantum fluctuations of the cavity and mechanical resonator modes. Equation~(\ref{Ae03}) can be separated to two different sets of equations, one for the mean values, and the other for the fluctuations, which are given by
\begin{eqnarray}\label{Ae04}
\dot{\alpha}&=&-i\left[\Delta_{c}^{\prime}+\xi\nu\cos(\nu t)\right]\alpha-\frac{\kappa}{2}\alpha-iE,\cr\cr
\dot{\beta}&=&-i\omega_{m}\beta-\frac{\gamma}{2}\beta+ig|\alpha|^{2},\cr\cr
\dot{\delta a}&=&-i\left[\Delta_{c}^{\prime}+\xi\nu\cos(\nu t)\right]\delta a-\frac{\kappa}{2}\delta a+iG(\delta b^{\dagger}+\delta b)
			-\sqrt{\kappa}a_{\mathrm{in}},\cr\cr
\dot{\delta b}&=&-i\omega_{m}\delta b-\frac{\gamma}{2}\delta b+iG\delta a^{\dagger}+iG^{\ast}\delta a
			-\sqrt{\gamma}b_{\mathrm{in}},
\end{eqnarray}
where $\Delta_{c}^{\prime}=\Delta_{c}-g(\beta^{\ast}+\beta)$ is the effective detuning modified by optomechanical coupling, $G=g\alpha$ is linearized optomechanical coupling strength, and we have neglected the nonlinear terms $ig\delta a(\delta b^{\dagger}+\delta b)$ 
and $ig\delta a^{\dagger}\delta a$ due the strong coherent driving conditions. The noise operations satisfy the following autocorrelation functions
\begin{eqnarray}\label{Ae05}
\langle a_{\mathrm{in}}(t)a_{\mathrm{in}}^{\dagger}(t^{\prime})\rangle&=&\delta(t-t^{\prime}),\cr\cr
\langle b_{\mathrm{in}}(t)b_{\mathrm{in}}^{\dagger}(t^{\prime})\rangle&=&(n_{\mathrm{th}}+1)\delta(t-t^{\prime}),
\end{eqnarray}
where $n_{\mathrm{th}}=\left\{\mathrm{exp}\left[\hbar\omega_{m}/(k_{B}T)\right]-1\right\}^{-1}$ is the mean thermal excitation number of the mechanical resonator at temperature $T$, $k_{B}$ is the Boltzmann constant. Then we can obtain the linearized Hamiltonian [see Eq.~(\ref{e02}) in the main text] in the presence of frequency modulation.

%\bibliographystyle{apsjnl}%
%\bibliography{BibtexRef2}
\end{document}